\documentclass[%
 reprint,
 amsmath,amssymb,
 aps,
]{revtex4-2}
\usepackage{tikz}
\usepackage{graphicx}
\usepackage{dcolumn}
\usepackage{bm}
\usepackage{color}

\usepackage{wasysym}
\usepackage{oplotsymbl}

\graphicspath{{figures/}{./}}

\begin{document}

\preprint{APS/123-QED}

\title{Collective behavior of self-steering active particles with velocity alignment and visual perception}
\thanks{Electronic supplement available }%

\author{Rajendra Singh Negi}
 \email{r.negi@fz-juelich.de}
\author{Roland G. Winkler}%
\author{Gerhard Gompper}
 \email{g.gompper@fz-juelich.de}
\affiliation{%
 Theoretical Physics of Living Matter, Institute of Biological Information Processing and Institute of Advanced Simulation, 
 Forschungszentrum J{\"u}lich, 52428 J{\"u}lich, Germany
}%

\date{\today}

\begin{abstract}

The formation and dynamics of swarms is wide spread in living systems, from bacterial bio-films to schools of
fish and flocks of birds. We study this emergent collective behavior in a model of active Brownian particles
with visual-perception-induced steering and alignment interactions through agent-based simulations. 
The dynamics, shape, and internal structure of the emergent aggregates, clusters, and swarms of these intelligent
active Brownian particles (iABPs) is determined by the maneuverabilities $\Omega_v$ and $\Omega_a$, quantifying 
the steering based on the visual signal and polar alignment, respectively, the propulsion velocity, characterized
by the P{\'e}clet number $Pe$, the vision angle $\theta$, and the orientational noise. Various non-equilibrium 
dynamical aggregates  -- like motile worm-like swarms and millings, and close-packed or dispersed clusters -- are obtained.  
Small vision angles imply the formation of small clusters, while large vision angles lead to more complex clusters.
In particular, a strong polar-alignment maneuverability $\Omega_a$ favors elongated worm-like swarms, which display 
super-diffusive motion over a much longer time range than individual ABPs, whereas a strong vision-based maneuverability 
$\Omega_v$ favors compact, nearly immobile aggregates.  Swarm trajectories show long persistent
directed motion, interrupted by sharp turns. Milling rings, where a worm-like swarm bites its own tail,  emerge for 
an intermediate regime of $Pe$ and vision angles. Our results offer new insights into the behavior of animal swarms, 
and provide design criteria for swarming microbots.

\end{abstract}

\maketitle


\section{Introduction}
\label{sec:intro}

Self-organized group formation and collective motion in form of swarms or flocks is 
a hallmark of living systems over a wide range of length scales, from bacterial biofilms 
to school of fish, flocks of birds, and animal herds 
\cite{elgeti_2015_RPP,bechinger_2016_RMP,ramaswamy_2010_ARCMP, shaebani_2020_NRP, vicsek_2012_PhysRep,cavagna_2014_ARCMP}. 
This behavior emerges without central control and is rather governed by the response of individuals to 
the action of other group members or agents. The self-organized structures and motion typically 
extend over much larger length scales than the size of the individual units, and emergent properties and function 
achieved are beyond the capacity of constituent units 
\cite{pearce_2014_PNAS, partridge1982structure, Gomez-Nava_2022_Nature, aoki_simulation_1982,  ballerini_2008_PNAS}.
Arising patters and structures not only depend on the physical
interactions between the various agents of an ensemble, but are often governed by 
nonreciprocal information input, e.g., visual perception in case of animals,
processing of this information, and active response. Unravelling the underlying mechanisms 
and principles not only sheds light onto the behavior of biological systems, but provides 
concepts to design functional synthetic active \cite{bechinger_2016_RMP, dai_2016_NN} and microrobotic 
\cite{rubenstein_programmable_2014} systems, which are able to adopt to environmental 
conditions and perform complex tasks autonomously \cite{tsang_2020_AIS, hartl_2021_PNAS}.

Several interactions and information-exchange processes can contribute to formation of swarms and flocks. 
Various models have been proposed and analyzed to understand this process and the resulting structures. 

A pertinent feature of the collective motion of animal groups is motion alignment and cohesion. 
In a pioneering work, Reynolds proposed three interaction rules for the bird-like objects called ``boids", 
which are collision avoidance, velocity matching, and flock centering \cite{reynolds_1987_POTACCGIT}. 
The boid model shares feature with earlier models on fish, which considered alignment with nearby individuals, 
attraction to the center of the school, and avoidance of close neighbors to prevent collisions 
\cite{aoki_simulation_1982,partridge1982structure}.
A similar model is the ``behavioural zonal model" by Couzin et al. \cite{couzin_2002_Elsevier, couzin_2005_nature}, 
which considers different types of interactions in three non-overlapping zones: minimum distance, attraction
towards other individuals, and alignment with neighbours.
The analysis of this model shows complex structures like swarms, millings, and groups with highly parallel motion.
From the physics perspective, perhaps the most celebrated model of collective motion is the ``Vicsek model"  
\cite{Vicsek_1995_PRL}, and its refinements and extensions 
\cite{toner_1995_PRL,Aitor_2018_soft_matter,vicsek_2012_PhysRep, gregoire_2004_PRL, szabo_2006_PRE, zhao_2021_PRE, chate_2008_PRE, kursten_2020_PRL}. 
In the basic version of this model, particles move with constant speed and change their direction at each 
time step by aligning with the mean orientation of neighboring particles in a prescribed interaction range, 
together with some noise accounting for environmental perturbations. The Vicsek 
model shows a phase transition from disordered phase to order phase with increasing density and decreasing noise. 

Another class of models emphasizes the short-range steric repulsion, and possibly longer-range attraction between the 
self-propelled units, called ``active Brownian particles" (ABPs).
This implies that the shape of the objects is now relevant to determine structure formation and collective motion. 
Motility-induced phase separation is observed in such systems, where uniformly distributed spherical ABPs can phase 
separate into the dense phase of slow-moving particles and dilute fast-moving particles under certain packing fractions 
and activity \cite{Claudio_2020_PRL, solon_2015_PRL, pasquale_2018_PRL}, while ABPs with elongated shapes form 
non-equilibrium motile clusters and swarm \cite{peruani_2006_PRE, abkenar_2013_PRE,bar_2020_ARCMP}. 

Models just based on an attraction of individual units to the center of mass of neighbouring
particles induced by self-steering controlled by visual perception 
\cite{strombom2011collective, barberis_2016_PRL, Renaud_2020_Science, negi_2022_soft_matter} show different
non-equilibrium structures like clusters, single-file motion, and millings. 
Millings have also been observed in simulations of other models \cite{Cheng_2016_NJP, pearce_2014_PNAS}. 
Using vision-based velocity alignment with time delays, agents can spontaneously condense into 'droplets' \cite{durve2018active} and increasing the activity and/or delay time of an active particle's attraction to a target point can induce a dynamic chiral state \cite{wang_2023_NC}.
The combination of the two mechanism of attraction and 
Vicsek-type alignment \cite{li_2019_PRE} 
yields a shift of the critical noise amplitude of the phase transition, and type of phase transition, 
compared to the case of pure velocity alignment \cite{Vicsek_1995_PRL}.
Detailed observations of flocks of surf scoters have also been used to infer individual interaction forces 
in the behavioural zonal model \cite{lukeman_2010_PNAS}.

After the recognition that in starling flocks a typical individual significantly interacts only with 7 or 8 closest 
neighbours \cite{ballerini_2008_PNAS}, models with metric-free, ``topological" interactions have also been 
studied \cite{ginelli_2010_PRL, peshkov_2012_PRL,kolpas_swarm_models_2013, rahmani_2021_CP}. 
For instance, the ``topological Vicsek model", in which particles align their velocity with 
neighbors defined through the first Voronoi shell, shows qualitative different results, like no
density segregation, compared to its metric counterpart. In Delaunay-based models
\cite{kolpas_swarm_models_2013}, the communication topology of the swarm is determined by Delaunay triangulation, 
where the rules of attraction and repulsion between neighboring individuals are 
the same as for the zone-based models, except that the region of attraction is unbounded. The 
results suggest that Delaunay-based models are more appropriate for swarms that are larger in 
number and more spatially spread out, whereas the zone-based models are more appropriate for small groups.

We focus here on the numerical investigation of a minimal cognitive flocking metric model with 
visual perception \cite{barberis_2016_PRL, negi_2022_soft_matter} and polar-based alignment
\cite{Vicsek_1995_PRL} for a system of self-steering particles. The usual ABPs are additionally
equipped with visual perception and polar-alignment interactions. 
The visual signal allows these ``intelligent ABPs" (iABPs) to detect the instantaneous position 
of center of mass of neighbouring particles within vision cone (VC), whereas polar alignment 
favors reorientation toward the average orientation of neighbors. Our model shares some basic features 
with the behavioral zonal 
model \cite{couzin_2002_Elsevier}, like long-range attraction, short-range repulsion, and medium-range alignment. 
However, it is important to note that in our study, we incorporate (i) hard-core excluded-volume interactions 
instead of a zone of repulsion between particles \cite{couzin_2002_Elsevier}, or even point particles 
\cite{barberis_2016_PRL, li_2019_PRE}, and (ii) a limited maneuverability in response to external
signals. Additionally, the vision-based attraction in our model 
is non-additive, i.e. the reorientation force does not depend on the number of particles, but their (normalized)
distribution in the vision cone. A recent theoretical study \cite{chepizhko_2021_soft_matter} highlights the 
influence of the non-additive versus additive interactions on structure formation in the Vicsek model.
Experimentally, systems of intelligent ABPs can be realised by active colloids, which are steered externally 
by a laser beam, with an input signal mimicking visual perception 
\cite{Tobias_2020_Nature, Tobias_2020_Nature_Qurom_sensing, Francois_2019_Science}. 

The main goal of our study is the exploration of the state diagram -- which depends on several parameters, such as
propulsion, maneuverabilities for vision-induced steering and alignment, vision angle, and ranges of vision
and alignment interactions -- as well as a characterization of the emerging structures and dynamical behaviors.
We find various types of 
emergent structures like dispersed clusters, compact aggregates, worm-like swarms, and millings, resulting 
from the interplay of visual-signal-controlled steering and polar alignment. An important feature is the formation
of worm-like swarms with a large variability of elongation and thickness. The dynamics of the swarms displays
a persistent super-diffusive motion over a wide range of time scales, which becomes diffusive at long times.
This motion is characterized by a persistence length, controlled by the maneuverabilities and vision
angle. Furthermore, swarms are found to display interesting trajectories, with long periods of 
directed motion interrupted by sharp turns or circular arcs.


\section{Model} 
\label{sec:model} 

We consider a two-dimensional system of $N$ responsive ``intelligent" active Brownian particles (iABPs) with 
the position $\bm{r_i}(t)$ of particle $i$ ($i=1,\ldots,N$)at time $t$. The particles are self-steered with constant 
propulsion force $\bm F^{a}_i(t) = \gamma v_0 \bm e_i(t)$ along the direction $\bm e_i(t)$ and velocity $v_0$. The 
dynamics of this system is governed by the equations of motion \cite{das_2018_NJP, negi_2022_soft_matter}
\begin{equation} \label{eq:1}
m \ddot{{\bm r}_i} = -\gamma \dot{{\bm r}_i}   + \gamma v_0 \bm{e}_i + \bm F_i +  \bm{\varGamma}_i(t).
\end{equation}
Here, $m$ is the mass of an iABP, $\gamma$ the translational friction coefficient, $\bm F_i$ the force due to 
excluded-volume interactions between the iABPs, and $\bm \Gamma_i$ is a stochastic Gaussian and Markovian process of zero 
mean and the second moments $\bm\varGamma_i(t) \cdot \bm\varGamma_j(t') = 4 k_BT \gamma \delta(t-t') \delta_{ij}$, 
with $T$ the temperature and $k_B$ the Boltzmann constant. Excluded volume interactions are taken into account by the 
short-range, truncated, and shifted Lennard-Jones potential
\begin{align}
     U(r)= \begin{cases}  
          \displaystyle  4 \epsilon \left( \left(\frac{\sigma}{r} \right)^{12} - 
          \left(\frac{\sigma}{r} \right)^{6}\right) + \epsilon, & r\leq 2^{1/6} \sigma \\ \mbox{0,} & \mbox{otherwise} 
          \end{cases} ,
\end{align}
where $r = |{\bm r}|$ is the distance between iABP particles, $\sigma$ represents their diameter, and $\epsilon$
is the energy determining the strength of repulsion.   

\begin{figure}
	\includegraphics[width=.48\textwidth]{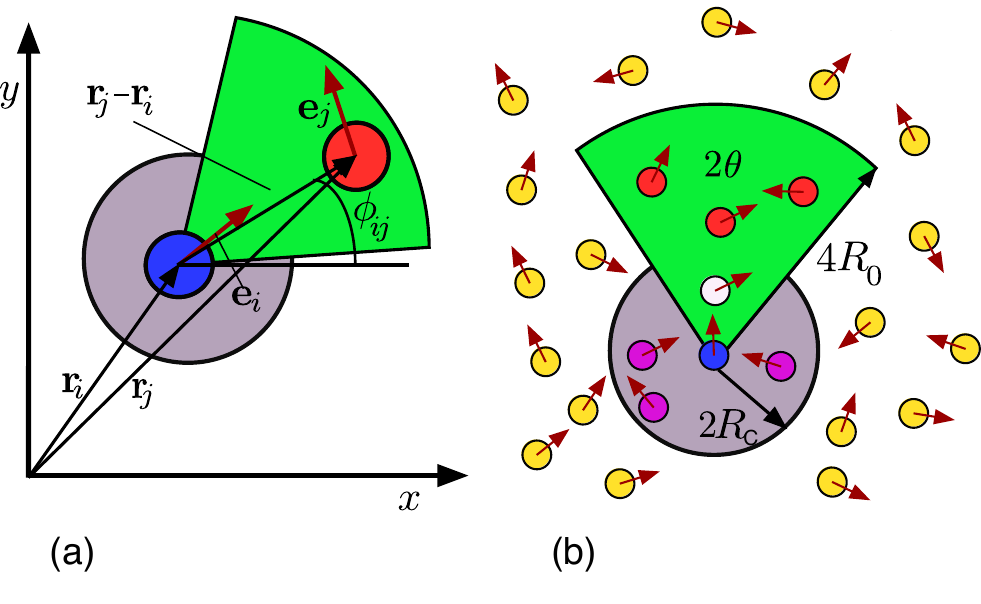}
	\caption{(a) Schematic representation of vision cone and alignment neighborhood of particle $i$ at position $\bm{r_i}$,
	with orientation $\bm{e_i}$, distance vector $\bm{r_j} - \bm{r_i}$ to other particles, and the corresponding orientation
	angle $\phi_{ij}$. 
	(b) Schematic showing the polar orientation field with cutoff $R_C$ and vision cone of blue particle with vision angle 
	$\theta$ and cutoff radius $4R_0$. 
	The blue particle interacts with other particles (red)
	through visual perception only within the vision cone (green), and aligns with other particles (pink) within the alignment
	region (grey).
	Particles (white) in the overlap region experience both interactions. 
	}
	\label{Graphical_representaion}
\end{figure}

An iABP is able to react to information about the position and orientation of neighboring particles.
As shown schematically in Fig.~\ref{Graphical_representaion}, particle $i$ at position $\bm{r_i} $ can adjust
its propulsion direction $\bm{e_i}$ through self-steering in the direction $\bm u_{ij} =({\bm r}_j - {\bm r}_i)/|{\bm r}_j - {\bm r}_i|$, 
determined by the positions of neighbors, with an adaptive force ${\bm f}^{av}_i$, as in the cognitive flocking model \cite{barberis_2016_PRL,negi_2022_soft_matter,Gomez-Nava_2022_Nature}. Simultaneously, it is capable to align its propulsion 
direction with those of neighboring particles,  $\bm{e_j}$, with the alignment force ${\bm f}^{aa}_i$, as in the Vicsek model \cite{vicsek_2012_PhysRep,Aitor_2018_soft_matter,chepizhko_2021_soft_matter}. Hence, the dynamics of the propulsion direction 
of particle $i$ is determined by \cite{goh_2022_NJP}
\begin{equation}\label{eq:adaptive_force}
    \dot{\bm{e}_i}(t) = \bm{f}^{av}_i + \bm{f}^{aa}_i + \bm\varLambda_i(t) \times \bm{e}_i(t) .
\end{equation}
Here, the $\bm\varLambda_i$ represent Gaussian and Markovian stochastic processes with zero mean and the correlations 
$\langle \bm\varLambda_i(t) \cdot \bm\varLambda_j (t') \rangle = 2D_R \delta_{ij} \delta(t-t')$, with the rotational 
diffusion coefficient $D_R$. 

The cognitive force (``visual'' force) is given by
\begin{equation}
     \bm{f}^{av}_i = \frac{\Omega_v}{N_{c,i}} \sum_{j\in VC}e^{-r_{ij}/R_0} \bm{e}_i \times ( \bm{u}_{ij}\times \bm{e}_i),
\end{equation}
with the ``visual" maneuverability $\Omega_v$ and the number  
\begin{align} \label{eq:eq_2_norm}
  N_{c,i} =  \sum_{j\in VC}e^{-r_{ij}/R_0} 
\end{align}
of iABPs within the vision cone (VC). The condition for particles $j$ to lie within the vision cone of particle $i$ is 
\begin{equation}
	\bm{u}_{ij} \cdot \bm e_i  \geq \cos(\theta) ,
\end{equation}
where $\theta$ --- denoted as vision angle in the following --- is the opening angle of the vision 
cone centered by the particle orientation $\bm e_i$ (Fig.~\ref{Graphical_representaion}). The exponential distance dependence 
describes a characteristic range $R_0$ of visual perception. In addition, we limit the vision range to  
$|\bm r_i - \bm r_j| \leq 4 R_0$ and treat all particles further apart as invisible.

Alignment of the propulsion direction (velocity alignment) is described by the adaptive force
\begin{equation}
    \bm{f}^{aa}_i = \frac{\Omega_a}{N_{a,i}} \sum_{j\in PA}\bm{e}_i \times ( \bm{e}_j \times \bm{e}_i),
\end{equation}
with the ``alignment" maneuverability $\Omega_a$, and the number $N_{a,i}$ of iABPs in the polar-alignment
circle (PA) (Fig.~\ref{Graphical_representaion}). The condition for particles $j$ to lie within the polar alignment (PA) range 
of particle $i$ is $|\bm r_i - \bm r_j| \leq 2 R_c$, where $R_c$ is the cutoff radius. Unless stated otherwise, $R_c=\sigma$, 
i.e., particles align up to second shell of neighbors. Particles inside the overlap zone of the VC and PA regions interact 
both via $\bm f^{av}_i$ and $\bm f^{aa}_i$.

The representation of the propulsion directions in polar coordinates,  
$\bm e_i = (\cos \varphi_i , \sin \varphi_i)^T$ (see Fig.~\ref{Graphical_representaion}), yields 
the equations of motion for the orientation angles $\varphi_i$,
\begin{equation} \label{eq:2}
\begin{aligned}
     \dot{\varphi_i} & = \frac{\Omega_v}{N_{c,i}} \sum_{j\in VC}e^{-r_{ij}/R_0} \sin({\phi_{ij}-\varphi_i})\\
      & +\frac{\Omega_a}{N_{a,i}} \sum_{j\in PA} \sin({\varphi_{j}-\varphi_i}) +  \Lambda_i(t) ,
  \end{aligned}
\end{equation}
The first sum on the right-hand side of Eq.~\eqref{eq:2} describes the preference of an iABP to move 
toward the center of mass of iABPs in its ``vision'' cone (VC), while the second sum describes the 
preference of an iABP to align with the neighboring particles. 
In the vision-based self-steering, the sum corresponds to the projection of the positions of all 
$N_c$ particles within the VC  onto the ``retina'' of particle $i$, with $\phi_{ij}$ the polar angle 
of the unit vector $\bm{u}_{ij} = (\cos \phi_{ij}, \sin \phi_{ij})^T$ between the positions of 
particles $i$ and $j$. The activity of the iABPs is 
characterized by the P\'eclet number
\begin{equation}
     Pe=\frac{\sigma v_0}{D_T} , 
\end{equation} 
where $D_T = k_B T/\gamma$ is the translational diffusion coefficient.


Our model presents a minimalistic description of self-steering particles with visual perceptions and 
velocity alignment, and provides insight into the interplay of these two swarming models.  However, it depends 
on a significant number of parameters, as there are the P{\'e}clet number $Pe$, the vision angle $\theta$ and 
the vision range $R_0$, the visual maneuverability $\Omega_v$, the alignment maneuverability $\Omega_v$, the 
particle size $\sigma$, and the packing fraction $\Phi$. In order not to get lost in this large parameter space,
we focus here varying the alignment-vision ratio $\Omega_a/\Omega_v$, the P{\'e}clet number $Pe$, 
and the vision angle $\theta$.


\section{Parameters}
\label{sec:params} 

In the simulations, we measure time in units of $\tau = \sqrt{m\sigma^2/(k_BT)}$, energies in units 
of the thermal energy $k_BT$, and lengths in units of $\sigma$. We choose 
$\gamma = 10^2 \sqrt{m k_BT/\sigma^2}$ and the rotational diffusion coefficient 
$D_R= 8 \times 10^{-2}/ \tau$, which yields the relation $D_T/(\sigma^2 D_R) = 1/8$. 
The above choice of the friction and rotational diffusion coefficients ensures that inertia does not 
affect the behavior, because the resulting relation $m/\gamma = 10^{-2} \tau \ll \tau$ implies 
strongly overdamped single-particle dynamics. 
The main reason for including the inertia term in Eq.~(\ref{eq:1}) is the reduced numerical effort 
and the improved accuracy of the numerical integration of Eq.~(\ref{eq:1}), as purely Brownian Dynamics
requires an orders of magnitude smaller time step.  We set $\epsilon/k_BT=(1+Pe)$ to ensure a 
nearly constant iABP overlap upon collisions, even at high activities. 
The iABP density is measured in terms of the global packing fraction $\Phi = \pi \sigma^2 N/(4L^2)$, 
with $L$ the length of the quadratic simulation box. Periodic boundary conditions are applied, 
and the equations of motion \eqref{eq:1} are solved with a  velocity-Verlet-type  algorithm suitable 
for stochastic systems,  with the time step $\Delta t = 10^{-3} \tau$ \cite{gronbech_2013_MP}. We perform $10^7$ 
equilibration steps, and collect data for additional $10^7$ steps. For certain averages, 
up to $10$ independent realizations are considered. As shown in Ref.~\cite{Suvendu_2019_PRL}, 
for the ratio $M= m D_R/\gamma = 8 \times 10^{-4}$ and the considered P\'eclet numbers, we do not 
expect MIPS. 

If not indicated otherwise, the number of particles is $N=625$, the length of the simulation box 
is $L=250\sigma$ -- corresponding to a packing fraction $\Phi= 7.85 \times 10^{-3}$ --, the 
characteristic radius is $R_0 = 1.5\sigma$, $\Omega_v/D_R=12.5$, and the range of vision angle is 
$\pi/16 \leq \theta \leq \pi$. 

Initially, the iABPs are typically arranged on a square lattice, with iABPs distances equal to 
their diameter $\sigma$, in the center of the simulation box.
In order to study the importance of vision and alignment in the interplay between these two 
self-steering mechanisms, we keep the vision-based maneuverability $\Omega_v/D_R$ constant, and 
vary the alignment-vision ratio by changing the alignment-based maneuverability $\Omega_a$.

\begin{figure*}
    \includegraphics[width=.98\textwidth]{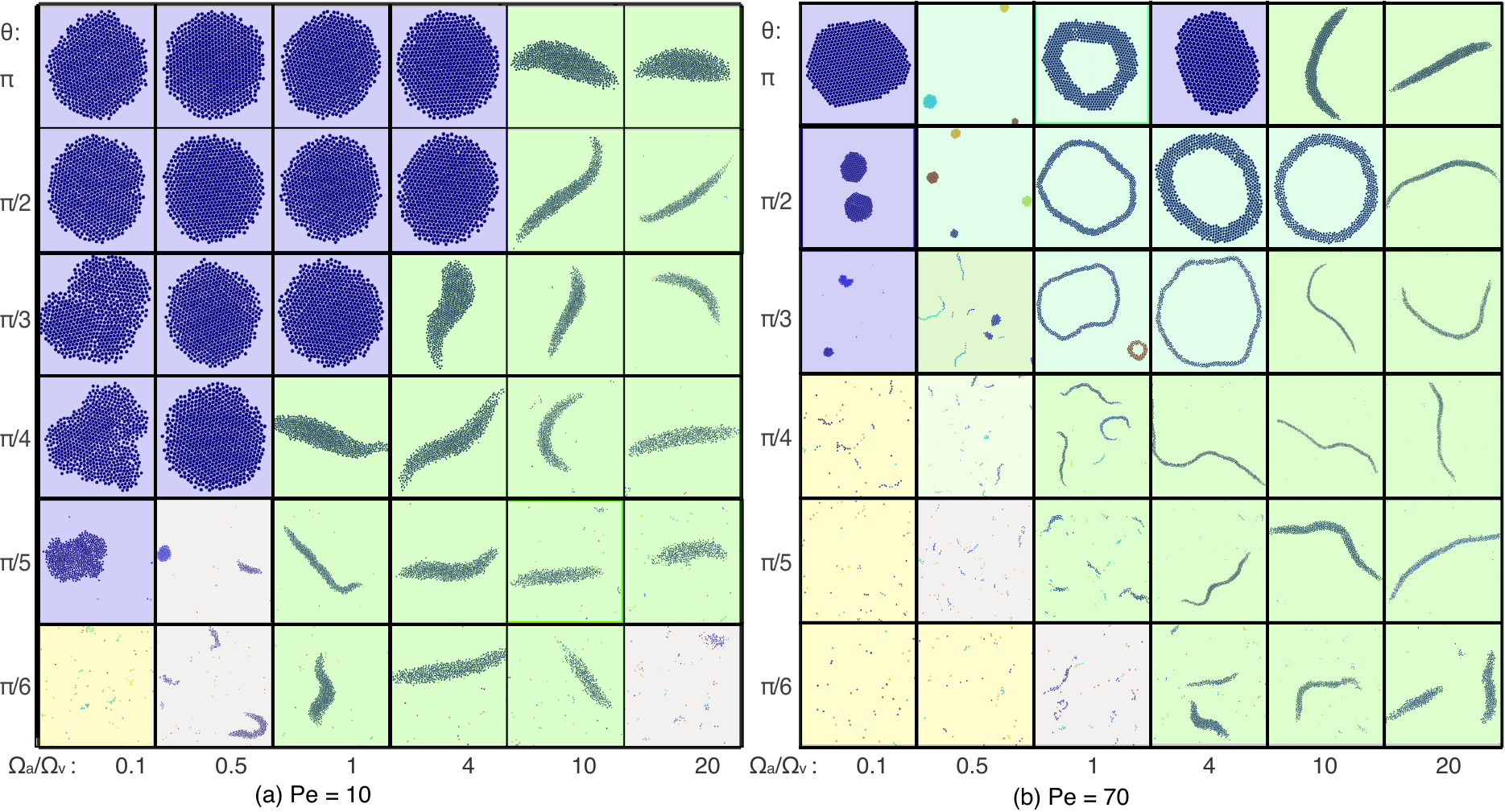}
    \caption{Snapshots of emerging structures for various vision angles $\theta$, alignment-vision 
    ratios $\Omega_a / \Omega_v$, and the P\'eclet numbers (a) $Pe=10$ and (b) $Pe=70$. In order to ensure 
    clear visibility, the snapshots are not presented to scale. For certain structures, a zoomed-in view is 
    necessary to provide a more detailed representation, see SM Fig.~S1 for scaled structures. The dilute phase is highlighted within yellow box, 
	dispersed clusters are represented in grey box, close packed cluster in purple box, and worm-like swarms in green box.}
    \label{fig:snapshot_various_vision-alignmnet_ratio}
\end{figure*}


\section{Results}
\label{sec:structural_dynamical}

\subsection{Phases and phase diagram}
\label{sec:phase_diagram}

\subsubsection{Phase behavior -- vision-induced steering versus polar alignment}
\label{sec:general}

The effect of the maneuverability ratio of polar alignment and vision-induced steering, $\Omega_a / \Omega_v$, and of 
the vision angle on the emerging structures is illustrated in Fig.~\ref{fig:snapshot_various_vision-alignmnet_ratio} 
for two P{\'e}clet numbers. The packing fraction $\Phi =0.00785$ is very low, hence, typically only a single or very 
few clusters or aggregates can be observed at any moment in time.

For the low P{\'e}clet number $Pe=10$, where orientational noise plays a significant role, the
state diagram in Fig.~\ref{fig:snapshot_various_vision-alignmnet_ratio}(a) shows two clearly distinguishable regimes, 
(i) the pursuit-dominated regime at low $\Omega_a / \Omega_v \lesssim 4$ and large vision angles $\theta \gtrsim \pi/4$, 
characterized by large quasi-circular, nearly immobile clusters, and 
(ii) the alignment-dominated regime at higher $\Omega_a/\Omega_v \gtrsim 10$ and smaller 
vision angles $\theta \lesssim \pi/5$, characterized by thick elongated worm-like swarms, 
which are highly mobile.

When alignment interactions dominate, $\Omega_a/\Omega_v \gtrsim 10$, the iABPs obviously tend to align 
in the same direction, but cohesion by vision-based steering toward clusters of other iABPs 
is still relevant; together, these two effects result in the formation of worm-like motile swarms for 
large vision angle $\theta=\pi$. As the vision angle is reduced to $\pi/5$, the number of particles 
in the vision cone decreases, thus, cohesion weakens, and the worm-like swarms become thinner 
and more elongated. An important point to note is that even for very small vision angles, 
i.e., $ \theta \leq \pi/8$, vision-based cohesion remains important for aggregate formation due to 
the very low packing fraction.

When the vision-mediated interaction dominates, i.e., for $\Omega_a / \Omega_v = 0.1$ and $0.5$, 
close-packed structures are observed for the vision angle $\theta \geq \pi/3$, and dilute 
structures for the lower angle $\theta \leq \pi/6$. These cases are similar to those of systems with
purely vision-based interactions \cite{negi_2022_soft_matter}. 
For large vision angles, a significant number of neighbors are sensed by an iABP, which then moves 
toward their center of mass easily, the effect of the alignment interaction is too weak to generate 
any significant parallel orientation and the iABPs form large close-packed aggregates. When the 
vision angle is low, e.g., $\theta =\pi/6$, 
very few particles are detected within the vision cone, no clusters can form, and particles are
distributed homogeneously.

For intermediate values $\Omega_a/ \Omega_v = 4$, close-packed structures are observe at the high vision 
angle $\theta= \pi$, while thick worm-like motile swarms emerge at the lower value $\theta =\pi/3$ (see also Supplementary Movies M1 and M2). 

The effect of vision-based steering becomes weaker with decreasing vision angle, as fewer particles 
appear in the vision cone. This can be captures by an effective maneuverability 
$\Omega_{v,eff} = \Omega_v \theta^\nu$ with $\nu \ge 1$. We will show below that $\nu \simeq 2$.
Thus, vision-base steering dominates at large vision angles, which favors compact clusters, 
and alignment for intermediate vision angles, which favors worm-like swarms. The wiggling of the 
worm-like swarm arises from the orientational noise of the leading particles.

For the higher P{\'e}clet number $Pe=70$, where persistent ballistic motion becomes more prevalent,
the characteristic emergent structures are displayed in Fig.~\ref{fig:snapshot_various_vision-alignmnet_ratio}(b). 
At high alignment-vision ratio, again elongated worm-like swarms are observe, which are, however, 
much thinner compared 
to those of the lower activity case with $Pe=10$. A new feature is the
emergence of milling structures, where thin worm-like swarms ``bite their own tail" and form ring-like
rotating aggregates; they are observed for $1 \le \Omega_a/ \Omega_v \le 10$, and vision angles 
$\pi/3 \le \theta \le \pi/2$ (see Supplementary Movie M3 ).  
In the vision-dominated regime, with $\Omega_a/ \Omega_v= 0.5$, we observe small rotating clusters or a coexistence-phase with 
small worm-like swarms and small rotating aggregates at the vision angle $\theta \ge \pi/3$ (see Supplementary Movies M4, M5) . A phase of 
small worm-like swarms is found for $\theta \le \pi/4$, which is similar to worm-aggregate 
phase and single-file motion in the system without alignment interactions \cite{negi_2022_soft_matter}, 
except that the aggregates are here rotating and are smaller in size.
We like to emphasize that we use different initial conditions for all parameter 
sets in order to avoid a bias by the initial condition toward some rare configuration, in
particular, for the milling structures. The highly elongated worm-like 
swarms can sometimes show milling intermittently, but then regain the worm-like conformation 
(see Supplementary Movie M6). Yet, the milling conformations displayed in
Fig.~\ref{fig:snapshot_various_vision-alignmnet_ratio}(b) always remain stable over the whole
simulation time. For close-packed structures at $Pe=10$ in 
Fig.~\ref{fig:snapshot_various_vision-alignmnet_ratio}(a), we employ an initial
configuration, where particles are distributed uniformly. This leads first to the formation of 
multiple close-packed aggregates, which subsequently merge to form a single large cluster. 
A different behavior is observed for the small rotating aggregates at $Pe=70$, 
e.g., at $\Omega_a/ \Omega_v= 0.5$ and $\theta \ge \pi/3$ in Fig.~\ref{fig:snapshot_various_vision-alignmnet_ratio}(b), 
which do not merge, but rather form by splitting of an initial large aggregate in the center of the 
simulation box. Thus, the small rotating clusters at high activities are different from the large 
close-packed aggregates observed at lower activities.     


\subsubsection{Phase Behavior -- Alignment-Dominated Regime}
\label{sec:alignment_dominated}

\begin{figure}
	\includegraphics[width=.46\textwidth]{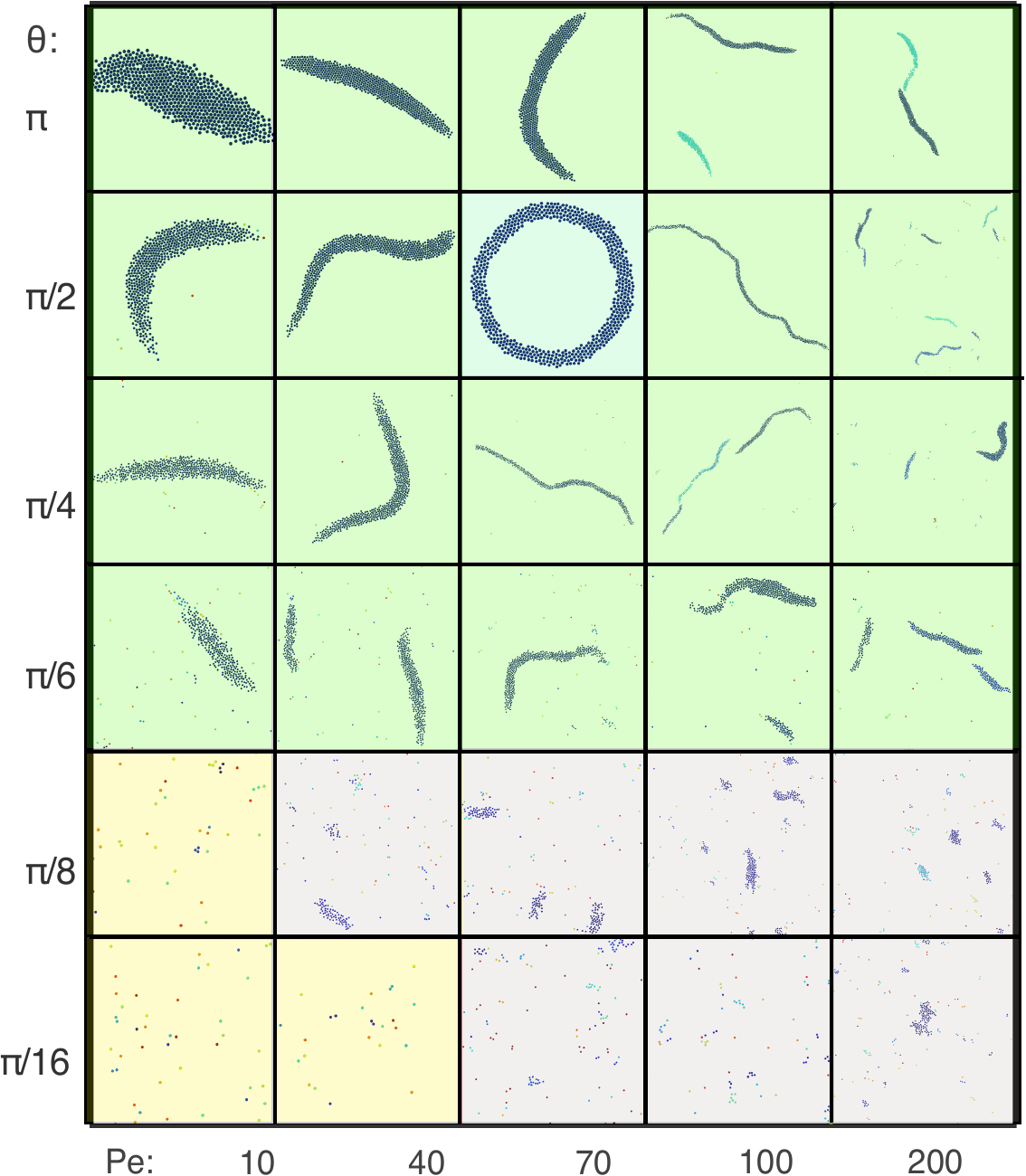}
	\caption{Snapshots of iABP structures are presented for various vision angles $\theta$, Péclet numbers $Pe$, 
	and an alignment-vision ratio $\Omega_a / \Omega_v = 10$.  The snapshots are not to scale for better visualization. See SM, Fig. S2 for full phase diagram.}
	\label{fig:Snapshot_1}
\end{figure}

For a more detailed investigation of the alignment-dominated regime, we focus on the alignment-vision ratio 
$\Omega_a/\Omega_v=10$. This provides insight into the structural evolution with increasing activity, characterized 
by the P{\'e}clet number $Pe$ and the vision angle $\theta$. Figure~\ref{fig:Snapshot_1} shows typical snapshot 
of emerging structures, like thin and thick worm-like swarms, milling, dispersed clusters, and a dilute phase as 
a function of these two parameters. For large vision angles $\theta \ge \pi/4$, predominately long and thick motile 
worm-like swarms are present. For vision angles $\theta \le \pi/6$, either dilute or dispersed clusters dominate. 

With increasing propulsion, 
the large worm-like swarms become thinner and more elongated as long as $\theta \gtrsim \pi/4$, while 
for $\theta \le \pi/8$  small aggregates persist. At high activity, $Pe \ge 100$, the large swarms show 
dynamical splitting into multiple swarms, while small swarms can merge into larger swarms. The very thin 
worm-like swarms can sometimes span the  whole system. There is a small window of parameters ($Pe\simeq 70$, $\theta\simeq \pi/2$) where 
circular milling-like structures appear. 


\subsubsection{Phase Behavior -- Balanced Alignment-Vision Regime}
\label{sec:balanced}

\begin{figure}
\includegraphics[width=.48\textwidth]{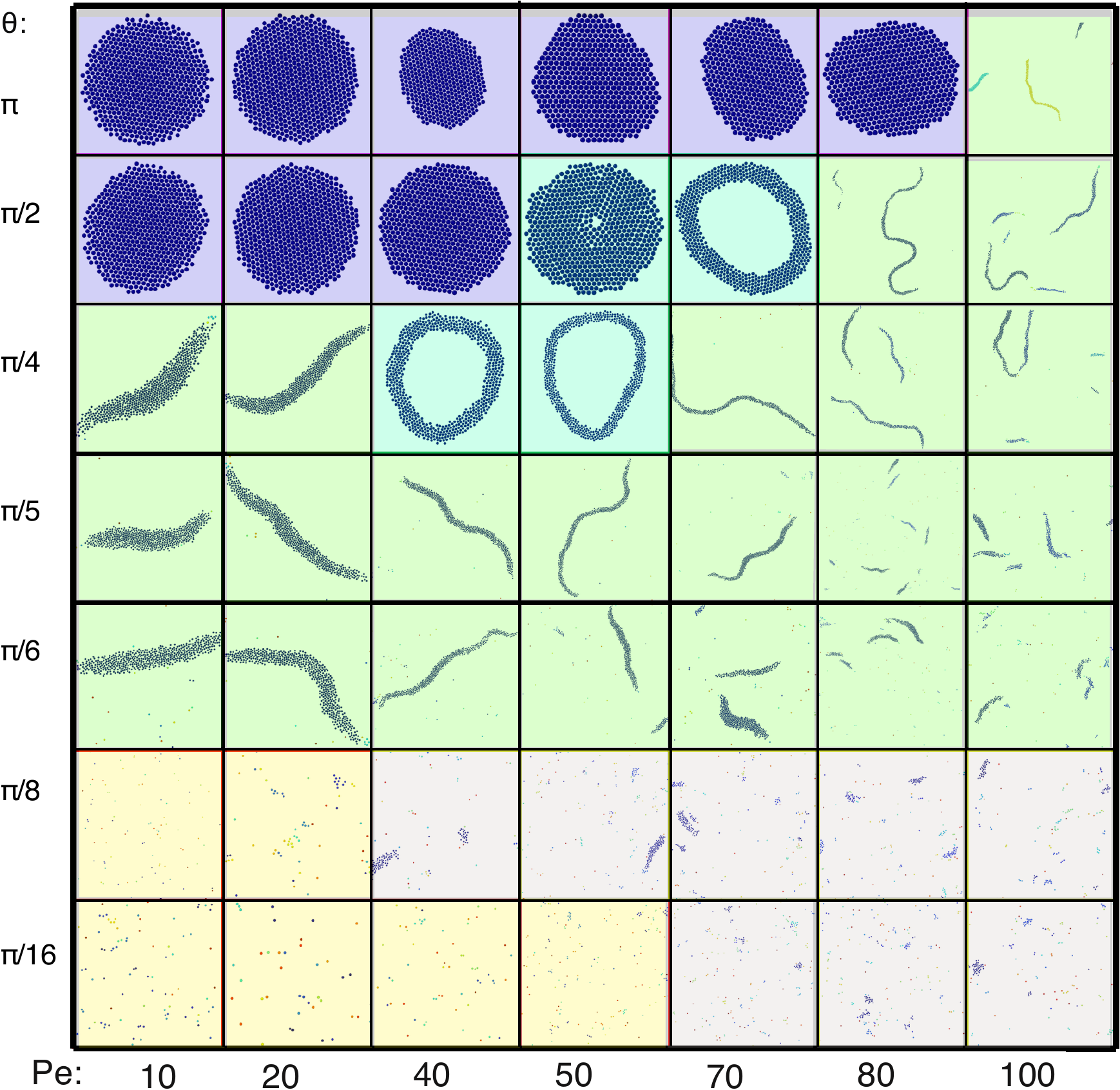}
	\caption{Snapshots of emerging structures for different activities $Pe$, vision angles 
	$\theta$, and alignment-vision ratio $\Omega_a / \Omega_v = 4$ for the packing fraction $\phi=0.00785$.
	The snapshots are not to scale for better visualization. }
	\label{fig:Phase_diagram_1}
\end{figure}

Figure~\ref{fig:snapshot_various_vision-alignmnet_ratio} indicates that $\Omega_a/ \Omega_v =4$ 
marks roughly the boundary between stationary close-packed compact structures -- where 
vision-based attraction dominates -- and motile worm-like swarms -- where alignment interactions
dominate. Thus, these two types of interactions approximately balance each other at this
alignment-vision ratio.

Snapshot of typical emerging structures at different $Pe$ and vision angles $\theta$ are 
displayed in Fig.~\ref{fig:Phase_diagram_1}. For vision angles $\theta \le \pi/8$, either 
dilute or dispersed cluster are obtain across all activities. With increasing vision angle 
$\theta \ge \pi/6$, first worm-like swarms, then compact clusters are stabilized.
For higher activity, $Pe \ge 100$, the close-packed structures are absent even at the maximum 
possible vision angle $\theta=\pi$, because the turning radius of a particle, determined by
$ Pe/(\Omega_v/D_R)$ \cite{goh_2022_NJP},
becomes too large for fixed maneuverability at high $Pe$ to reach the target cluster.
There is a transition from close-packed clusters to thick elongated worm-like swarms at 
$Pe=10$ and $20$ when the vision angle decreases from $\theta=\pi$ to $\pi/6$, similar as
in Fig.~\ref{fig:snapshot_various_vision-alignmnet_ratio}(a). 
At intermediate activities, $Pe=40$ to $70$, and intermediate vision angles
$\pi/2 \ge \theta \ge \pi/4$, milling structures appear, which are also present at 
$\Omega_a/\Omega_v=1$ and $10$ for $Pe=70$ 
(compare Fig.~\ref{fig:snapshot_various_vision-alignmnet_ratio}(b)). 

\begin{figure}
	\includegraphics[width=.48\textwidth]{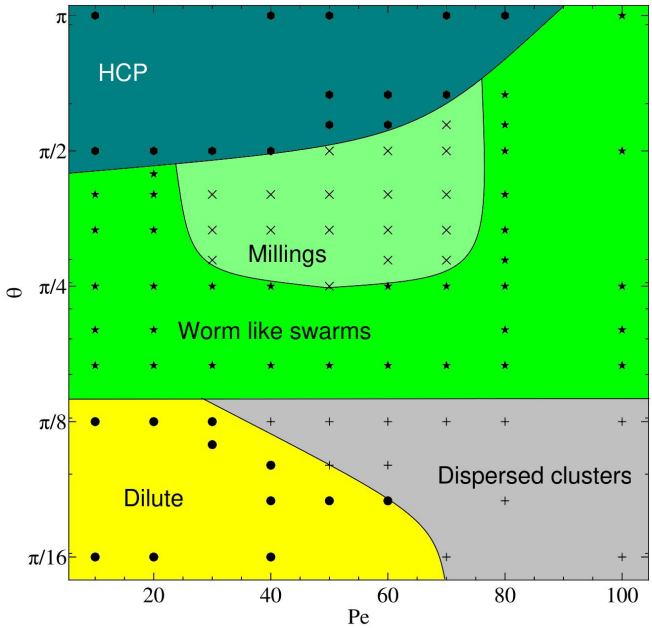}
	\caption{$Pe-\theta $ phase diagram for $N=625$, the ratio $\Omega_a/\Omega_v=4$, the  visual 
	    maneuverability $\Omega_v/ D_R=12.5$, and the packing fraction $\phi=0.00785$. 
		The individual phases are indicated by different colors and symbols:  
		HCP: navy ${\hexagofill}$,  worm: green {$\bigstar$}, dilute: yellow ${\CIRCLE}$. 
		The lines at the ``phase'' boundaries are guides to the eye. }
	\label{fig:Phase_diagram_ratio_0_25_system_size}
\end{figure}

The full phase diagram is shown in Fig.~\ref{fig:Phase_diagram_ratio_0_25_system_size}, 
where the boundaries between the various phases are clearly delineated. The comparison of 
Figs.\ref{fig:Phase_diagram_1} and \ref{fig:Phase_diagram_ratio_0_25_system_size} 
shows a remarkable feature of the milling structures; as the boundary to the 
close-packed clusters is approached, the milling band fills more and more in the interior, 
and at $\theta=\pi/2$ and $Pe=50$ becomes a milling disc.

\subsubsection{Discussion}

Figures~\ref{fig:snapshot_various_vision-alignmnet_ratio}, \ref{fig:Snapshot_1}, \ref{fig:Phase_diagram_1},
and \ref{fig:Phase_diagram_ratio_0_25_system_size} together provide a overview of the emerging structures 
in the three-dimensional parameter space of $Pe$, $\theta$, and $\Omega_a/\Omega_v$.

The main characteristics are the presence of (i) compact clusters in the vision-induced steering regime, with
$\Omega_a/\Omega_{v,eff} \le 4$, where alignment plays a minor role, (ii) worm-like swarms
in the alignment-dominated regime, with $\Omega_a/\Omega_{v,eff} \ge 10$, where the elongation of the 
swarm increases and the thickness decreases with increasing $Pe$ and increasing $\Omega_a/\Omega_{v,eff}$, 
and (iii) millings at intermediate values of $\Omega_a/\Omega_{v,eff}$ and $Pe$. 

The transition from close-packed aggregates to thick worm-like swarms occurs as the effective 
alignment-vision ratio increases from 
$\Omega_a/(\Omega_{v} \theta^\nu) \simeq 1$, both with increasing $\Omega_a$ 
(see Fig.~\ref{fig:snapshot_various_vision-alignmnet_ratio}) -- due to stronger 
alignment -- as well as decreasing vision angle (see Fig.~\ref{fig:Phase_diagram_1}) 
-- due to weaker vision-induced steering.

Long and thin worm-like swarms are favored by larger activities, due to a larger inward-pushing
force of particles at the swarm edges, as studied and explained in more detail in 
Sec.~\ref{sec:internal} below.
Long and thin worm-like swarms are also 
favored by small vision angles, as the thickness is related to the range $R_0 \theta$ of 
the vision cone. If the swarm thickness becomes larger than $R_0 \theta$, particles on the rim 
cannot see the full swarm width, and the swarm can split, similar to the single-file motion
observed in the vision-only case \cite{barberis_2016_PRL, negi_2022_soft_matter}.

Importantly, alignment stabilizes persistent swarm motion 
(compared to the single-file motion of vision-only systems \cite{negi_2022_soft_matter}), 
because the incipient leader particle becomes aware of and is affected by its followers.

It is important to note that the presence of worm-like swarms in our model at the low packing 
fraction ($\phi=0.00785$) is in stark contrast to the structures observed in the Vicsek model 
at higher packing fraction (e.g. $\Phi=0.25$), where homogeneous disorder phases and giant motile 
aggregates coexisting with a dilute gas of single particles are observed \cite{Aitor_2018_soft_matter}. Increasing the field of vision range yields a comparable outcome to enhancing the visual maneuverability of particles (see SM S-II). Similarly, extending the range of polar alignment demonstrates an effect akin to improving alignment-related maneuverability (see SM S-III).

\subsection{Structural Properties}


\subsubsection{Internal Structure of Worm-like Swarms}
\label{sec:internal}

\begin{figure}
    \centering
    \includegraphics[width=0.45\textwidth]{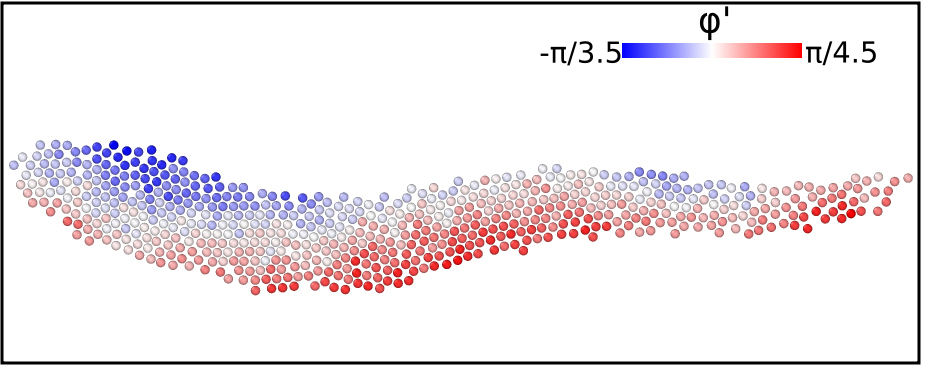}
    \caption{Snapshot of worm-like swarm where particles are colored according to the orientation with respect to average orientation  at $Pe=40$, $\Omega_a/\Omega_v=10$, and $\theta=\pi/2.$ }
    \label{fig:worm_oreint}
\end{figure}

An interesting feature of worm-like swarms is the increasing elongation and thinning  with increasing $\Omega_a$ and increasing $Pe$. This is related to the behavior of particles at the edge of the swarm, which, due to the 
vision-induced steering, push ``inwards", but, due to the strong alignment, can do so only to a limited extent. 
The balance of vision-induced steering and alignment forces can be employed in a simple mean-field estimate (see SM S-IV) to predict the particle 
orientation angle $\varphi^*$ at the edge of the swarm, with 
\begin{equation}
\varphi^* = \pm \theta \left[ 1+ \frac{\Omega_a}{\Omega_v} \frac{\theta \sin(\theta)}{1-\cos(\theta)} \right]^{-1}.
\end{equation}
This estimate is in semi-quantitative agreement with the orientational structure
of snapshots of worm-like swarms, see Fig.~\ref{fig:worm_oreint}. 

The preferred tilt angles $\varphi^*$ imply a lateral compressive force and equivalent
perpendicular P{\'e}clet number
\begin{equation}
    Pe_\perp = Pe \sin{\varphi^*}
\end{equation}
which is increasing with $Pe$, in agreement with the conformations
in Fig.~\ref{fig:Snapshot_1}. Furthermore,
the snapshot shows that there is an interesting correlation of particle orientation
and local curvature of the swarm centerline, where an imbalance of particles with inward
orientation on the two sides seems to generate the snake-like motion of the swarm.

\subsubsection{Swarm Shape and Asphericity}
\label{sec:asphericity}

We characterize the overall size and shape of the emerging structures by the 
radius-of-gyration tensor \cite{singh_2012_JOP}
\begin{equation}
  G_{mn} = \frac{1}{N} \sum_{i=1}^{N} \Delta r_{i,m} \Delta r_{i,n},
\end{equation} 
where $\Delta r_i$ is the distance of $i$-th particle from a cluster's center of mass,  
$m,n \in \{x,y\}$, and $N$ is the total number of particles in the cluster. We use a
distance criterion to define a cluster, where an iABP belongs to a cluster when its 
distance to another iABP is within a radius of $\sigma_0$. Since our system is 
very dilute, we choose $\sigma_0 =2\sigma$. In order to avoid averages to be strongly 
affected by configurations which occur only rarely, we only consider realizations 
which appear in more than $1\%$ of the recorded configurations. 

\begin{figure}
    \includegraphics[width=.45\textwidth]{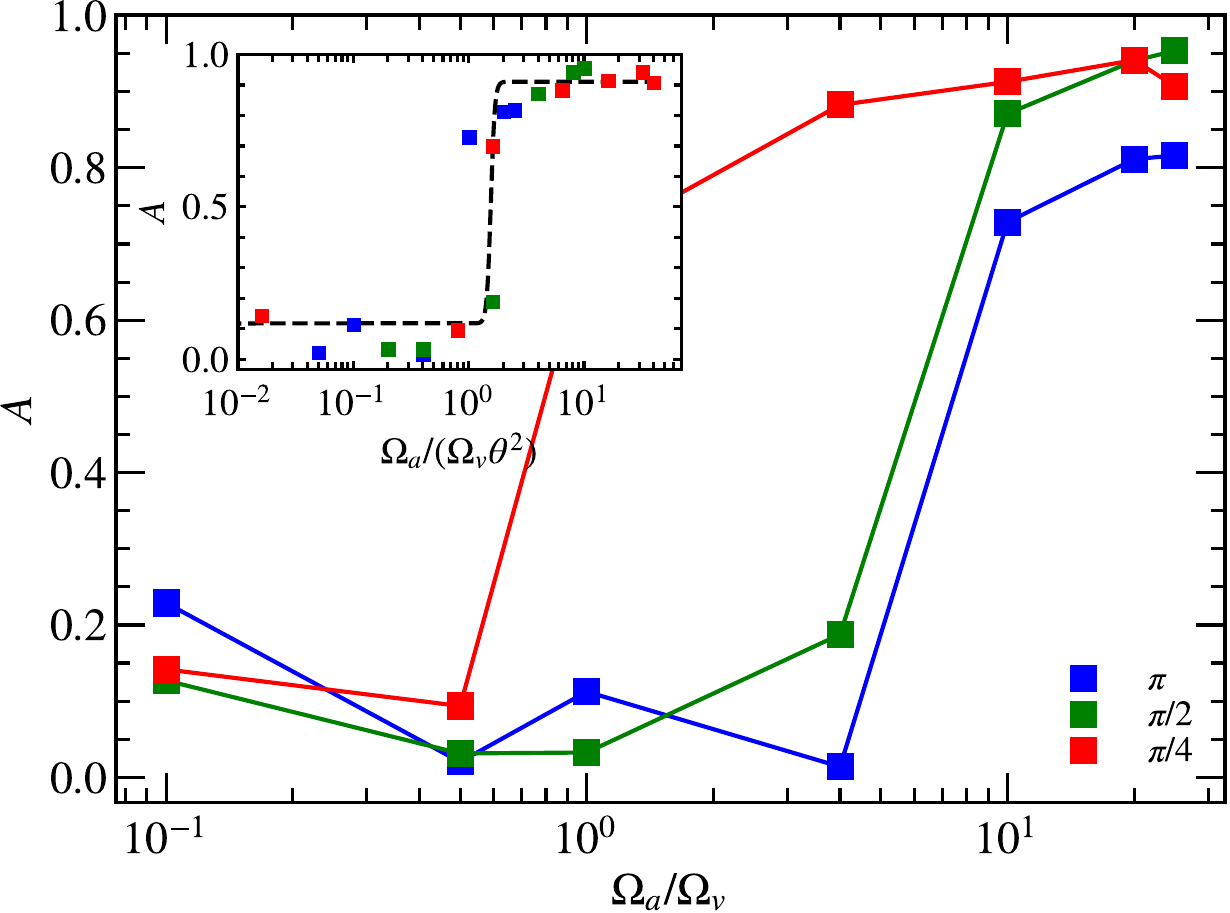}
    \caption{Aggregate asphericity $A$ as a function of the alignment-vision ratio $\Omega_a/\Omega_v$ 
    for $Pe=10$ and the indicated vision angles.}
    \label{fig:asphericity_as_function_of_ratio}
\end{figure}

An important quantity to characterize the shape of aggregates is the asphericity
\begin{equation}
    A= \frac{|\lambda_1-\lambda_2|}{\lambda_1+\lambda_2} ,
\end{equation}
where $\lambda_1$ and $\lambda_2$ are the eigenvalues of the radius-of-gyration tensor.
Figure~\ref{fig:asphericity_as_function_of_ratio} shows the asphericity $A$ as a
function of alignment-vision ratio at various vision angles $\theta$. The 
close-packed structures for weak alignment and strong vision at small 
$\Omega_a/\Omega_v \le  0.5$ are nearly circular, hence, $A\simeq 0$, similar 
to the vision-only case  \cite{negi_2022_soft_matter}. 
The worm-like swarms for $10 \le \Omega_a/ \Omega_v \le 25$ and $\theta=\pi$, as well 
as $\Omega_a/ \Omega_v=1$ for $\theta=\pi/4$, are highly elongated, 
which results in the large apshericities $A \simeq 0.8$. 
The asphericity starts to increase with $\Omega_a/\Omega_v$ significantly earlier for
smaller vision angles, because cohesion and thus formation of compact aggregates 
is suppressed for smaller visual signals, which favors worm-like swarms.

Thus, the effect of an increase of the vision angle is similar to an enhanced 
visual maneuverability, because in both cases the tendency of an iABP to steer 
toward existing clusters increases. Consequently, 
we re-calibrated the visual maneuverability $\Omega_v$ by a factor, which increases with the vision angle $\theta$ to 
accommodate this effect. As a result, we can collapse all data of the asphericity
for $\theta \leq \pi/4$ onto a single master curve by employing 
an effective scaled variable $\Omega_a/ (\Omega_v \theta^\nu)$, with $\nu \simeq 2$, 
as demonstrated in the inset of Fig.~\ref{fig:asphericity_as_function_of_ratio}. This shows that the asphericity 
displays universal behavior as a function of this scaled alignment-vision ratio, with a 
sharp transition from the compact-cluster to the worm-like swarm phase at 
$\Omega_a/ (\Omega_v \theta^2) \simeq 2$. 

A similar scaling behavior is found for the radius of gyration, see SM S-V.


\subsubsection{Global Polarization}
\label{sec:polarization}

\begin{figure}
	\includegraphics[width=.48\textwidth]{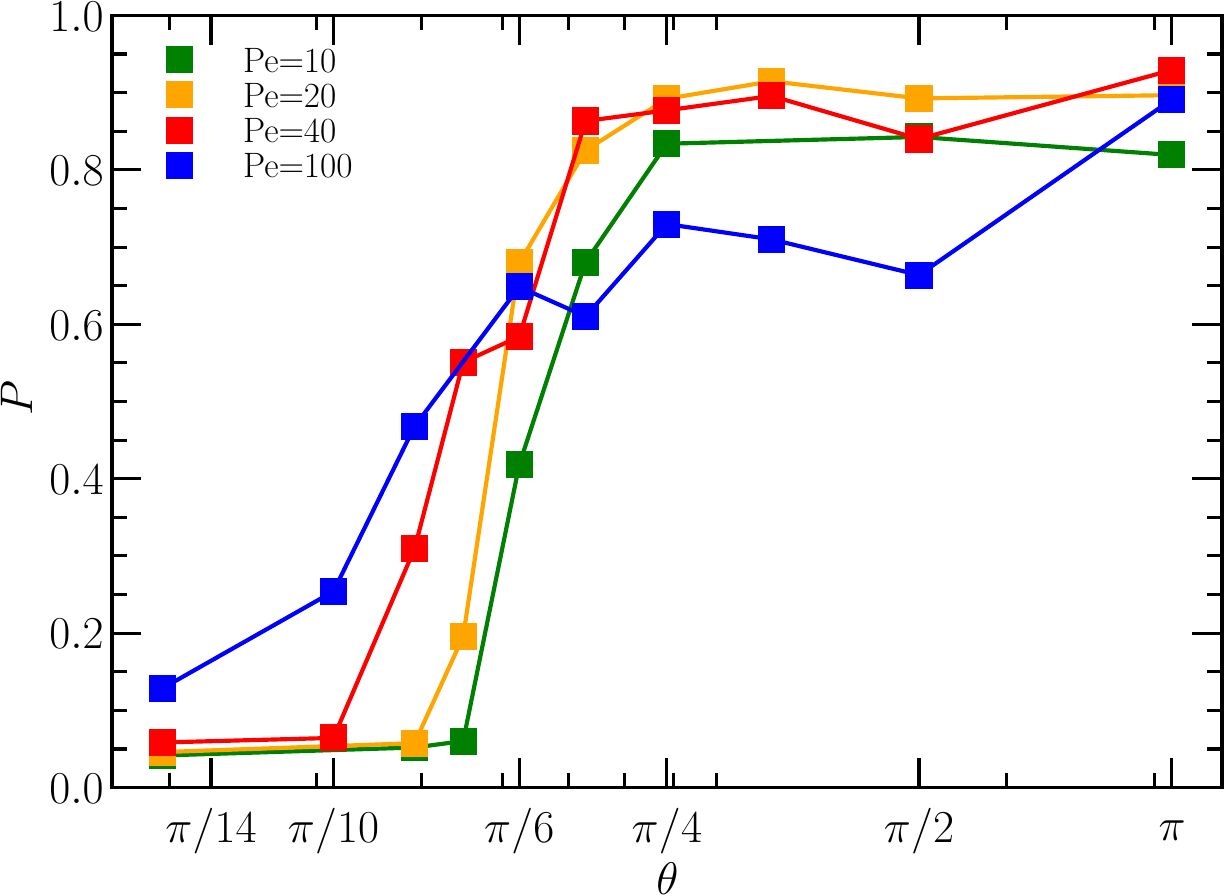}
	\caption{Polarization order parameter $P$ as a function of the vision angle $\theta$ for the indicated  
	activities $Pe$ and $\Omega_a / \Omega_v = 10$. The packing fraction is $\phi=0.00785$. }
	\label{fig:Polarization}
\end{figure}

The global polarization is characterized by the order parameter
\begin{equation} \label{eq:polarization}
	P=  \left\langle \frac{1}{N} \left| \sum_i \mathbf{e}_i \right| \right\rangle ,
\end{equation}
where ${\bm e}_i$ is orientation of particle $i$ and the average is performed over time. 
Figure~\ref{fig:Polarization} shows the polarization as a function of vision angle 
$\theta$ for $\Omega_v / \Omega_a = 10$ at various activities $Pe$. 
For $\theta \lesssim \pi/8$, particles are randomly oriented and $P\simeq 0$. For 
larger vision angles, global polarization emerges, which  can reach $P=1$
for $\theta = \pi$. Global polarization at small vision angles is enhanced 
by larger $Pe$, due to stronger persistence particle motion. It is important to 
note that we are not characterizing bulk phases here, but typically a single large 
cluster. Thus, $P$ quantifies the alignment order within the cluster. $P \simeq 1$ 
also does not imply that the cluster is always moving in the same direction, just 
that the propulsion directions of the individual particles remain highly aligned 
at any moment in time.

\subsection{Dynamical Properties}
\label{sec:dynamics}

\subsubsection{Mean-Square Displacement}
\label{sec:MSD}

The translational motion of the iABPs is characterized by their mean-square 
displacement (MSD)
\begin{equation}\label{eq:MSD}
    \langle \bm r^{2}(t) \rangle= \frac{1}{N} \sum_{i=1}^N \left\langle \left( \bm r_i(t+t_0)- \bm r_i(t_0) \right)^2 \right\rangle ,
\end{equation}
where the average is performed over the initial time $t_0$.
An important reference case is the behavior of single ABPs, for which theoretical calculations in two dimensions yield 
\begin{equation}\label{eq:MSD_2}
	\langle \bm r^{2}(t) \rangle =4D_T t + \frac{2v_0^2}{D_R^2} \left( D_R t -1 + e^{-D_R t}\right) ,
\end{equation}

\begin{figure}
    \centering
   \includegraphics[width=0.48\textwidth]{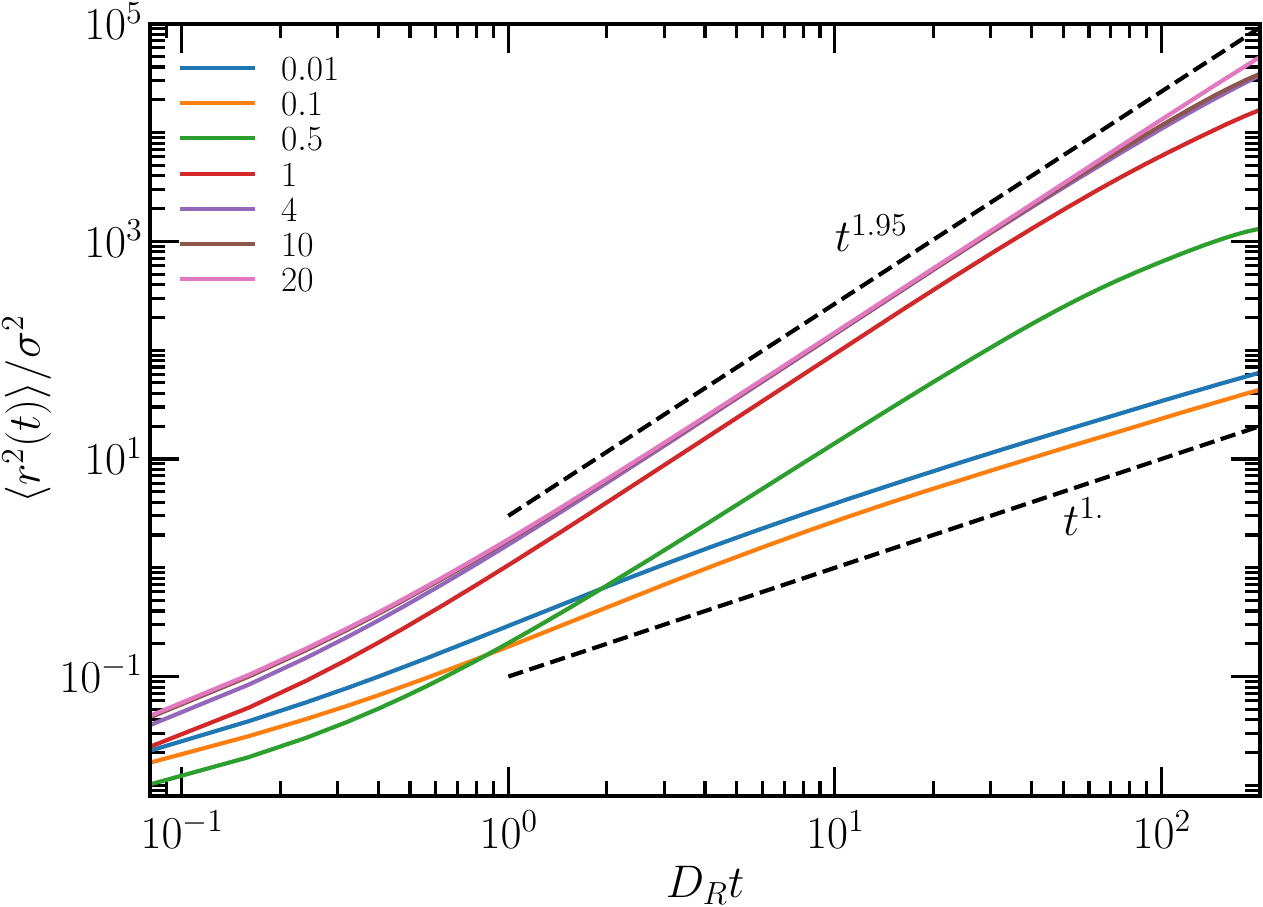}
   \caption{Mean-square displacement of iABPs as a function of time for $Pe=10$, $\theta=\pi/4$, and the various indicated ratios $\Omega_a/\Omega_v$.}
    \label{fig:MSD}
\end{figure}

Figure~\ref{fig:MSD} displays mean-square displacements of iABPs for 
various alignment-vision ratios, vision angles, and P{\'e}clet numbers. For larger 
$\Omega_a/\Omega_v=4$ to $25$, where  alignment interactions dominate over vision-controlled 
steering  (worm-like swarms), the particles move nearly ballistic and 
$\langle \bm r^{2}(t) \rangle \sim t^{\alpha}$, 
with the exponent $\alpha \approx 1.95$. For $\Omega_a/\Omega_v \le 0.1$, in the vision-dominated regime, the close-packed aggregates display 
translational diffusion and $\langle \bm r^{2}(t) \rangle \sim t$. The transition from ballistic
diffusive  motion occurs at $\Omega_a/ \Omega_v \simeq 0.5$ for 
$\theta=\pi/4$. It shifts to $\Omega_a/ \Omega_v \simeq 1$ for $\theta=\pi/2$ -- 
in agreement with the conclusion in Sec.~\ref{sec:asphericity} that the importance 
of vision-controlled steering increases with increasing vision angle.


\subsubsection{Collective Dynamics}

\begin{figure}
    \centering
    \includegraphics[width=.48\textwidth]{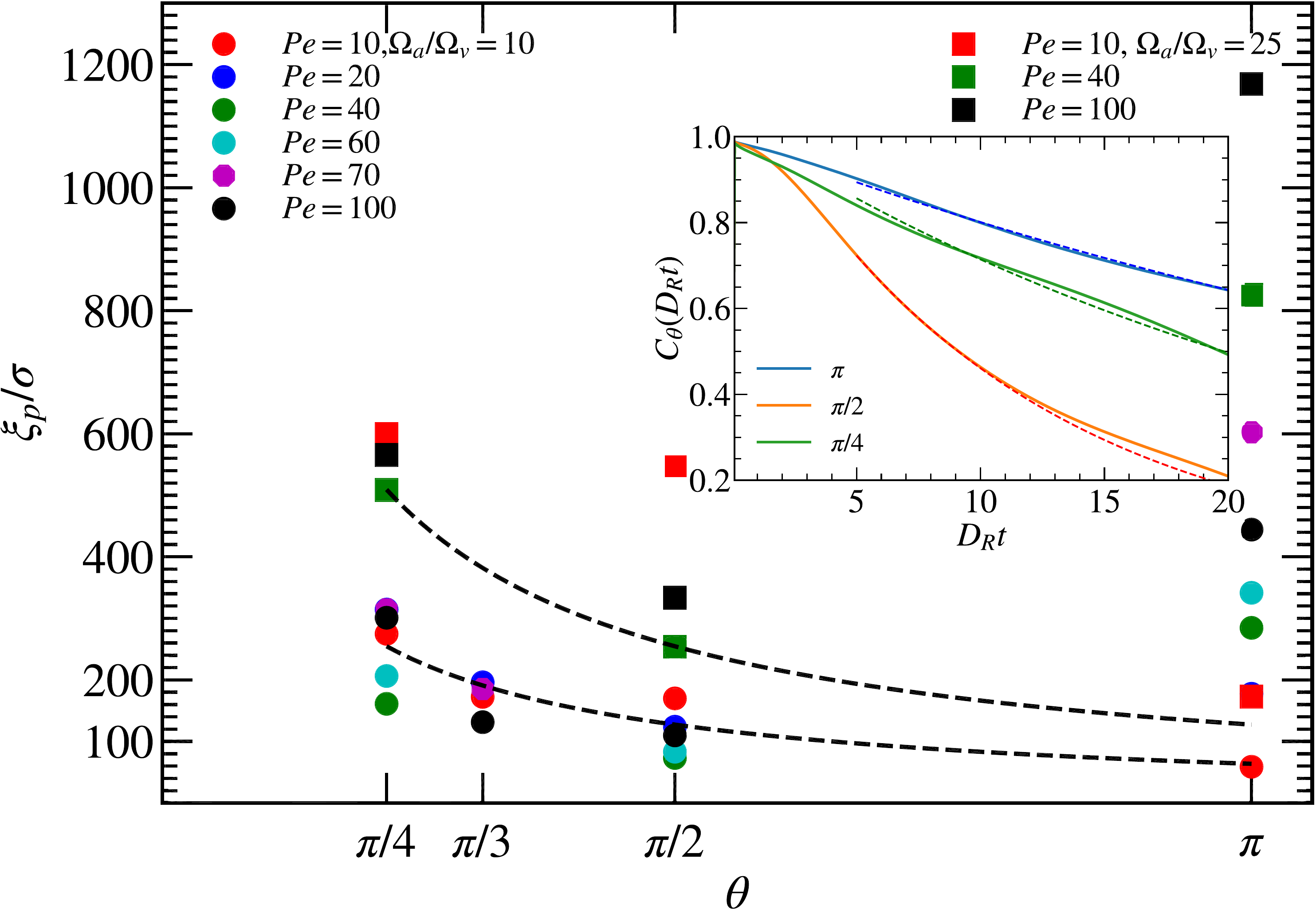}
    \caption{Persistence  length as function of vision angle $\theta$ across various $Pe$ and $\Omega_a/\Omega_v =10$, 
    represented with circles symbols, and $ \Omega_v/ \Omega_a=25$, represented in rectangle symbols. 
    Inset: auto correlation function at $Pe=60$ and $\Omega_v / \Omega_a =10$, dashed lines are fits.   }
    \label{fig:correlation length}
\end{figure}

The dynamics of elongated worm-like swarms is characterized by an essentially
one-dimensional motion along a curvi-linear path, where all particles of the swarm 
trace out trajectories, which are only slightly displaced laterally from the trajectory of the center-line (see Movie M7). This is reminiscent of the ``railway motion"
performed by semi-flexible, tangentially driven active polymers at high P{\'e}clet numbers \cite{isele_holder_2015_soft_matter}. 

Thus, we can characterize the dynamics of the whole swarm by the temporal auto-correlation function of individual particles,
\begin{equation}\label{eq:OCF}
     C_{\theta}(t) = \frac{1}{N} \sum_{i=1}^N \left\langle  {\bm e}_i(t+t_0) \cdot 
     {\bm e}_i(t_0) \right \rangle ,
\end{equation}
where, $\bm{e}_i$ represents the orientation of particle $i$, and $N$ is the total number of particles in the swarm. 

In the case of railway motion, the spatial conformations of the swarm as well as 
the temporal auto-correlation function, Eq.~\eqref{eq:OCF} are determined by 
the statistical properties of the (infinitely long) rail, with the spatial 
correlation function of tangent vectors ${\bf t}(s)$ (with contour length $s$)
\begin{equation}\label{eq:railway}
     \langle {\bf t}(s) \cdot {\bf t}(s') \rangle  = A \exp(-|s-s'|/\xi_p)
\end{equation}
and persistence length $\xi_p$. This length is also the spatial correlation length of 
shape fluctuations of the center-line of the swarm. Furthermore, the ``railway" assumption implies that
\begin{equation}\label{eq:railway1}
     \langle  {\bm e}_i(t+t_0) \cdot {\bm e}_i(t_0) \rangle \simeq \langle {\bf t}(s) \cdot {\bf t}(s+v_0t) \rangle  = A \exp[-(v_0/\xi_p)t]
\end{equation}
where $v_0$ is active velocity.
Thus, the temporal decay of $C_{\theta}(t)$ should be controlled by the 
relaxation time $\tau=\xi_p/v$, with the {\em same} persistence length $\xi_p$ as the
instantaneous conformations. 

Figure \ref{fig:correlation length} shows examples of the auto-correlation function
$C_{\theta}(t)$, together with exponential fits (inset), and the derived persistence
lengths for various parameter combinations. It is interesting to note that a 
comparison of the spatial and the temporal (Eq.~\eqref{eq:railway1}, Eq.~ S-5)
persistence length are in good quantitative agreement for elongated worm-like swarms
(see SM S-VIII, Fig.~S8).
The persistence length $\xi_p$ display three important trends: (i) The persistence 
grows roughly linearly with the alignment-induced maneuverability $\Omega_a$, 
(ii) is only weakly dependent on the P{\'e}clet number, and 
(iii) it decreases with the vision angle roughly as a power law $\theta^{-1}$ in the range $\pi/4 \le \theta \le \pi/2$. 
Together this implies
\begin{equation}
    \xi_p/\sigma \simeq \Omega_a/(\Omega_v \theta)
\end{equation}
The increase of the persistent swarm motion (compared to the single-file motion of
vision-only systems \cite{negi_2022_soft_matter}) with stronger alignment-induced
maneuverability can be traced back to the effect of the followers on the incipient
leader particle through the (isotropic) alignment interaction.
The larger persistence length at $\theta = \pi/4$ than at $\pi/2$ can be attributed 
to the larger worm length at $\theta = \pi/4$, where a more focused vision  
enables the particles to more easily follow the incipient leader.

An important point to note here is that the persistence for the considered parameter combinations is always large, with $\xi_p/\sigma > 100$. Since the effective translational diffusion coefficient for a random-walk-like motion with
Kuhn length $\xi_p$ is given by
\begin{equation}
    D_T^{eff} \simeq v_0 \xi_p ,
\end{equation}
this explains the large ballistic/superdiffusive regime in Fig.~\ref{eq:MSD}, 
because the crossover from the ballistic to the diffusion regime occurs at 
$D_R t^* \simeq D_R \xi_p / v_0$, which implies 
$D_R t^* \simeq  8 (\xi_p /\sigma)/ Pe \simeq 100$.

A notable exception in Fig.~\ref{fig:correlation length} is the vision angle
$\theta=\pi$. In this case, the persistence length increases roughly proportional 
to the P{\'e}clet number. The main difference to the case of smaller vision angles 
is that the worm-like swarms are here much thicker (and shorter), and exhibit a 
less persistent motion for small $Pe$. This implies that 
the rotational diffusion of the leading group of particles -- which is determined by $Pe$ -- now plays an important role.
Furthermore, with increasing $Pe$, the swarm thickness decreases (see Fig.~\ref{fig:Snapshot_1}), which also contributes to an increasing persistence length.


\subsubsection{Milling}
\label{sec:millings}

Milling structures are characterized by the angular frequency 
\begin{equation}
     \omega  =  \frac{1}{N_e} \frac{\sum_{i}(\bm{r_i}-\bm{r_{cm}}) \cdot \bm{v}}{\sum_{i}(\bm{r_i}-\bm{r_{cm}})^2} 
\end{equation}
 and the radius 
 \begin{equation}
     R  =  \frac{1}{N_e} \sum_{i}\sqrt{(\bm{r_i}-\bm{r_{cm}})^2} 
\end{equation}
of these aggregates, here $N_e$ is total number of particles in the milling structure and $\bm r_{cm}$ is center-of-mass position.

\begin{figure}
	\includegraphics[width=.48\textwidth]{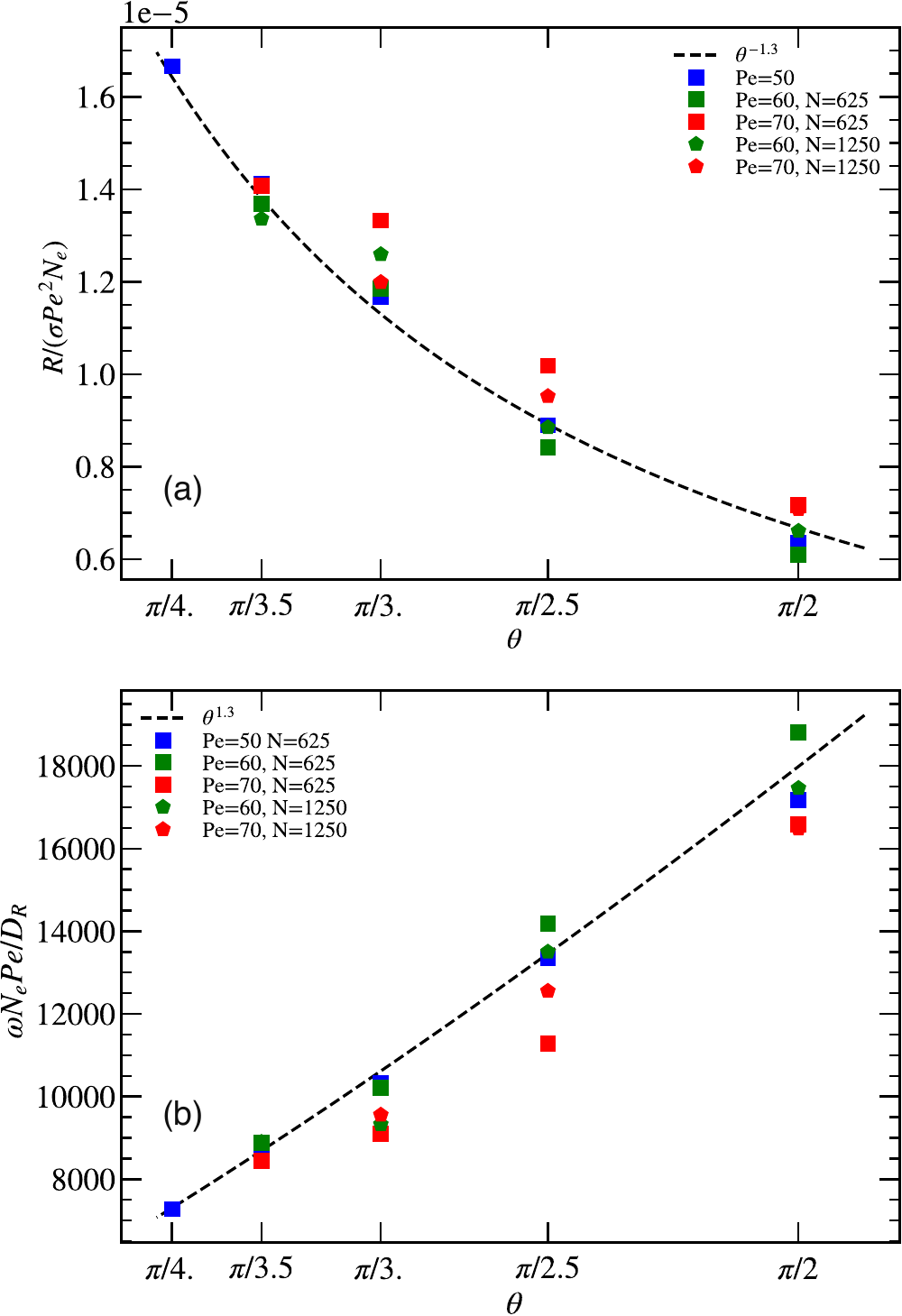}
	\caption{(a) Scaled radius $R$ and (b) scaled angular frequency $\omega$ of milling structure as 
	function of $\theta$ at alignment-vision ratio $\Omega_a/\Omega_v=4$, with different system size ($N$) of iABPs.}
	\label{fig:angular_frequency}
\end{figure}

The radius of the milling ring-like ribbon for $\Omega_a/\Omega_v=4$ and $\pi/4 < \Theta < \pi/2$, increases roughly as $R \sim Pe^2$ .
This is caused by the increasing persistence of the iABP motion with increasing $Pe$.   
Figure~\ref{fig:angular_frequency}(a)
shows the scaled radius as a function of the vision angle, and suggests that $R\sim \theta^{-\gamma}$, 
with $\gamma\simeq 1.3$. The angular frequency $\omega$ decreases with increasing activity $Pe$ 
approximately as $\omega \sim 1/Pe$. 
This decrease is related to the increasing radius, because $\omega \sim v/R$ and $v \sim Pe$.
Figure~\ref{fig:angular_frequency}(b) shows scaled frequency $\omega$ as a function of the vision angle, 
with $\omega \sim \theta^{\gamma}$ and the same exponent $\gamma$ as for the radius.
Thus, all together, we predict the scaling behavior
\begin{equation}
    R = c_R \sigma \theta^{-\gamma} N_e Pe^2, \ \ \ \ \omega = c_\omega \theta^{+\gamma} D_R/(N_e Pe)
    \label{eq:milling}
\end{equation}
with $\gamma\simeq 1.3$ and constants $c_R$ and $c_\omega$. The data in Fig.~\ref{fig:angular_frequency}
indeed fall very nicely onto these single scaling curves.



\subsection{Finite-Size Effects}
\label{sec:finite_size}

\begin{figure}
	\includegraphics[width=.48\textwidth]{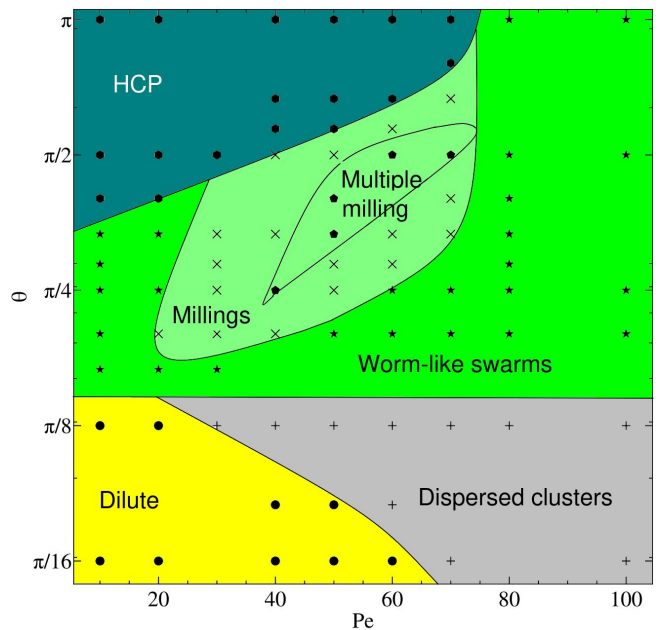}
	\caption{Phase diagram in $Pe$-$\theta$ space at the particle number 
	$N=1250$, the alignment-vision ratio $\Omega_a/\Omega_v=4.$, packing fraction
	$\phi=0.00785$, and vision-related maneuverability $\Omega_v/ D_R=12.5$. 
	The individual phases are indicated by different colors and symbols.  
	HCP: navy ${\hexagofill}$,  worm: green {$\bigstar$}, dilute: yellow ${\CIRCLE}$. 
	The lines at the ``phase'' boundaries are guides to the eye. }
	\label{fig:Phase_diagram_large_system_size}
\end{figure}

In order to elucidate the influence of finite-size effects on the presented results, we 
construct a phase diagram for the same parameters as in Fig.~\ref{fig:Phase_diagram_ratio_0_25_system_size},
for the alignment-vision ratio $\Omega_a/\Omega_v=4$, packing fraction $\phi=0.00785$, and vision
maneuverability $\Omega_v/ D_R=12.5$, but now of twice the number of particles, i.e., $N=1250$,
see Fig.~\ref{fig:Phase_diagram_large_system_size}. Overall, the topology of the phase diagram remains 
the same, mostly only phase boundaries are slightly shifted. 
The most significant change is the extension of the region of stability of the milling structure, which
extends to smaller $Pe$ numbers and smaller $\theta$ for larger $N$. 

This similarity does not imply that the iABP behavior is independent of $N$. An obvious effect of 
increasing $N$ is that the close-packed, nearly circular aggregates in the HCP phase grow in size, 
with their radius increasing as $\sqrt{N}$. As the particles are all pushing 
toward the joint center of mass, this implies that the aggregates become more 
stable, which is expressed by the shift of the HCP - worm-like swarms boundary 
to lower vision angles. For the worm-like swarms and the milling structures, 
these can either remain a single aggregate, or slit and merge again intermittently 
into several smaller structures. In the latter case, only minor finite-size 
effects can be expected. In the former case, increasing $N$ implies longer or 
thicker worm-like swarms or milling structures. Thicker swarms exhibit a more 
persistent and less "snaking-like" motion. This also appears for the milling 
structures, where $R\sim N$ and correspondingly $\omega \sim 1/N$, see
Fig.~\ref{fig:angular_frequency} and Eq.~(\ref{eq:milling}).

\section{Summary and Conclusions}
\label{sec:summary}

We have studied the emergent structures and dynamics of ensembles of cognitive, 
self-steering particles, with a combination of visual-perception controlled steering 
and polar alignment. The visual signal gives particles a tendency to reorient toward 
the center of mass of other particles in their visual field and implies group cohesion, 
whereas polar alignment induces particle reorientation toward the average orientation 
of their neighbors within the alignment-perception range and implies collective
directed motion. Depending on the vision-induced maneuverability $\Omega_v$ and 
polar-alignment related maneuverability $\Omega_a$, various kinds of collective motion 
are obtained. Moreover, the vision angle $\theta$, the vision range $R_0$, and the 
activity $Pe$ play a crucial role in structure formation. In worm-like swarms, 
which are predominately observed for large alignment-vision ratios 
$\Omega_a / \Omega_v \gtrsim 4$, particles move together collectively with little
individual orientational fluctuations, and 
the swarm display super-diffusive or nearly ballistic motion over long times. 
Dispersed cluster and dilute phases prevail at small vision angles, typically 
$\theta \leq \pi/8$, due to the small number of particles in the vision cone, 
which implies weak cohesion. Close-packed disk-like clusters emerge for high
vision-induced maneuverability and vision angle $\theta \geq \pi/4$, because the 
larger the number of particle in a visible cluster, the larger is also the tendency to quickly turn toward its center of mass. 

Circular milling structures are obtained mainly for balanced alignment-vision ratios, 
$\Omega_a / (\Omega_v \theta^2) \simeq 4$, at an intermediate range of activities and 
vision angles. The underlying mechanism for such structures to be stable is that the 
persistence length of the worm-like contour of a conformation should be on the same 
order as its contour length, because large persistence lengths favor elongated swarms.
The milling structure is characterized by a radius $R \sim Pe^2$ and, hence, a 
rotation frequency $\Omega \sim 1/Pe$. 
Balanced maneuverability, i.e., $\Omega_a / (\Omega_v \theta^2) \simeq 4$, seems to be
a very favorable condition for swarms in general, because it makes swarms susceptible 
to external perturbations while remaining cohesive, so that the swarm can quickly react 
to the appearance of predators. 
We want to mention parenthetically that the importance of critically in biological
systems has also been discussed in the context of scale-free correlation of swarms 
of midges \cite{attanasi2014prl}. 

A closer look at the internal structure of a worm-like swarm reveals interesting new
features. Particle orientations at the edge of the swarm are weakly inclined toward the
centerline, which implies a compressive force responsible for swarm elongation.
Furthermore, a lateral asymmetry seems to be correlated with undulations of the 
centerline.

Although our approach shares some basic features with the ``boid model" 
\cite{reynolds_1987_POTACCGIT} and the ``behavioural zonal model" 
\cite{couzin_2002_Elsevier}, it differs in other important aspects. 
First, we employ a hard-core repulsion between the agents, which implies close-packed 
aggregates, whereas the previous models typically adopt a softer repulsion potential, 
which leads to disordered aggregates. Thus, the way the repulsion between particles is 
modeled plays a crucial role in structure formation. It certainly depends on the 
real system to be considered, which of these repulsive interactions is more appropriate.
Second, while in the other models \cite{reynolds_1987_POTACCGIT, couzin_2002_Elsevier} 
attraction-related reorientation is instantaneous, 
in our model cohesion emerges from vision-based steering, 
where the reorientation toward a target is restricted by a limited 
maneuverability. Thus, both vision-  and alignment- related maneuverability are important 
parameters, which have not been investigated in combination with alignment so far.

\begin{figure}
    \centering
    \includegraphics[width=0.48\textwidth]{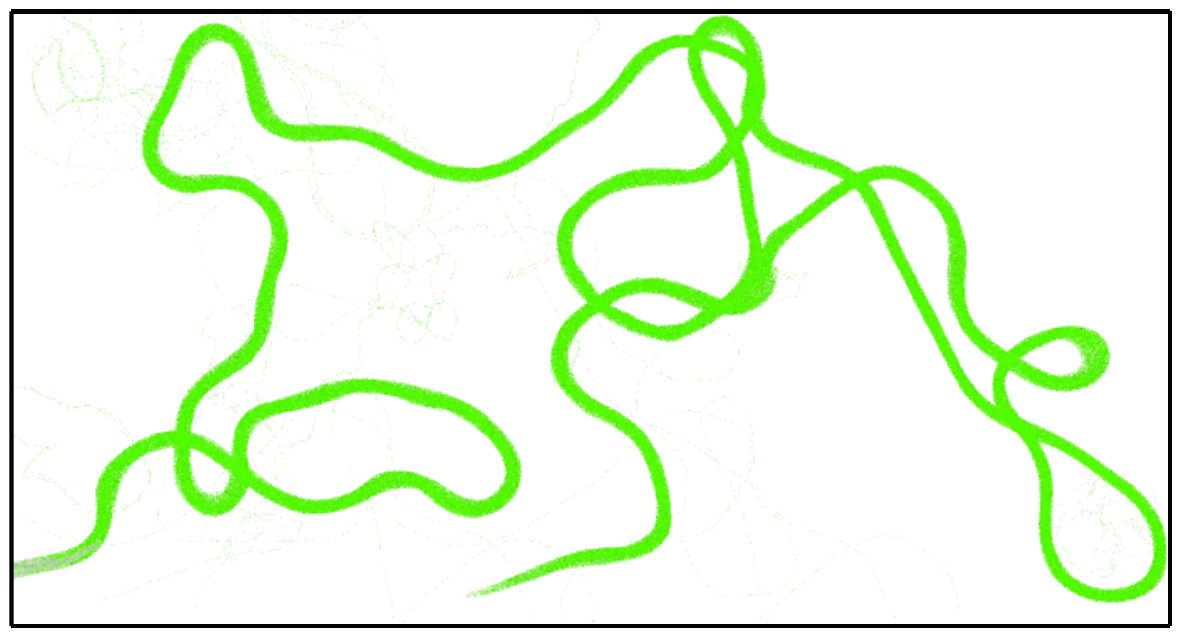}
    \caption{Typical trajectory of a worm-like elongated swarm at $Pe=40$, $\theta=\pi/2$ and $\Omega_a/\Omega_v=10$. }
    \label{fig:trajectory}
\end{figure}

Thick worm-like swarms have already been observed in the ``behavioral zonal model",
called highly parallel groups \cite{couzin_2002_Elsevier, couzin_2005_nature}. 
However, we also obtain highly elongated, thin worm-like swarms, in particular 
for large P\'eclet numbers and large alignment-vision ratios. These swarms have to be 
distinguished from the swarms in the pure vision-based minimal cognitive model with 
and without excluded volume \cite{barberis_2016_PRL, negi_2022_soft_matter}, where
they display single-file motion and are much shorter in length, i.e., less stable. 
The most interesting feature of these thin worm-like swarms is that they can transform 
into meta-stable milling states, where the swarm bites its own tail, and then regains 
the elongated shape later on. Milling structures have also been observed previously in 
the behavioural zonal model \cite{couzin_2002_Elsevier}. We observe both large 
(polar) milling bands and small rotating aggregates --- where the latter differ 
from the nematic ring-like bands in the vision-only case with point particles \cite{barberis_2016_PRL}. 

Millings have been seen previously in simulations of other models 
\cite{Cheng_2016_NJP, pearce_2014_PNAS}, but more importantly in groups of several
animal species in the wild, such as schooling fish \cite{Tunstroem_2012_fish}, 
army ants \cite{schneirla1944unique}, bats \cite{delcourt2016collective}, 
plant-animal worms \cite{franks_2016_PNAS, sendova2018plant}, and dictyostelium
\cite{rappel_1999_PRL}. Large extended worm-like swarms have been observed in 
flocks of birds \cite{cavagna_2014_ARCMP, Reynolds_2022_birdflocks}, herds of 
sheep, and school of fish.

We conclude from our simulations that it would be very interesting to study and
characterize the existence, motion, and trajectories of large worm-like swarms in 
more detail, both in simulations and in animals herd in the wild. We have analyzed 
the trajectories in terms of a persistent random walk model and
extracted effective persistence length. However, it is not at all obvious that 
the assumption of a persistent random walk fully captures the complexity of motion 
of an animal herd. In fact, a more detailed look at the long-time 
trajectory of a worm-like swarm, see Fig.~\ref{fig:trajectory}, already indicates 
that this behavior -- with long stretches of persistent directed motion interrupted by 
loop-like pieces and sharp turns  -- is much more complex and interesting than
a simple persistent random walk.





%

\end{document}